%% file: pi0-gamma-factorisation-breaking.tex
\documentclass[a4paper,11pt]{article}
\usepackage{pos}
\usepackage{subcaption}
\usepackage{slashed}

\input{useful-commands.tex}

\title{Breakdown of collinear factorisation in the photoproduction of a $  \pi ^{0}\gamma  $
	pair with large invariant mass}
\ShortTitle{Breakdown of collinear factorisation in the photoproduction of a $  \pi ^{0}\gamma  $
	pair}

\author*[a,b]{Saad Nabeebaccus}
\author[c,d]{Jakob Sch\"onleber}
\author[e]{Lech Szymanowski}
\author[a]{Samuel Wallon}

\affiliation[a]{Universit\'e Paris-Saclay, CNRS/IN2P3, IJCLab, 91405 Orsay, France}

\affiliation[b]{Department of Physics \& Astronomy, University of Manchester, Manchester M13 9PL, United Kingdom}

\affiliation[c]{Institut f\"ur Theoretische Physik, Universit\"at Regensburg, D-93040 Regensburg, Germany}

\affiliation[d]{RIKEN BNL Research Center, Brookhaven National Laboratory, Upton, NY 11973, USA}

\affiliation[e]{National Centre for Nuclear Research (NCBJ), 02-093 Warsaw, Poland}

\emailAdd{saad.nabeebaccus@manchester.ac.uk}
\emailAdd{jschoenle@bnl.gov}
\emailAdd{Lech.Szymanowski@ncbj.gov.pl}
\emailAdd{Samuel.Wallon@ijclab.in2p3.fr}

\abstract{We identify a $ 2 \to 3 $ exclusive process, where collinear factorisation is broken, namely the exclusive photoproduction of a $ \pi ^{0}\gamma $ pair with large invariant mass. This occurs because the process suffers from gluon exchanges trapped in the Glauber region. Using an explicit example, we show that the Glauber gluon, which is exchanged between a collinear spectator parton from the nucleon sector and a soft spectator parton from the outgoing pion, has both of its lightcone plus and minus components pinched. Therefore, it cannot be deformed to collinear/soft regions, as is often the case for processes that do factorise. We further confirm the leading power behaviour of the identified Glauber region, highlighting that this is the case although it relies on extracting a soft parton from the outgoing pion. We stress that the Glauber pinch for this process is of the leading power, due to the possibility of having two-gluon exchanges between the collinear nucleon sector and hard partonic scattering sub-process. In fact, the Glauber gluon that we identify is one of these two active gluons, and therefore, its effects are observed already at leading order. A direct consequence of our work is that collinear factorisation breaks in the same way for other $ 2 \to 3  $ exclusive processes, where two-gluon exchanges in the $ t $-channel are possible, like in the exclusive production of a photon pair from $  \pi ^{0} N $ collisions. However, we highlight that in cases where such two-gluon exchanges do not exist, like in the exclusive $   \pi ^{\pm}\gamma  $ photoproduction, the Glauber exchanges that we discuss here do not occur, and hence they do not suffer from factorisation breaking effects.}

\FullConference{31st International Workshop on Deep Inelastic Scattering (DIS2024)\\
 8–12 April 2024\\
Grenoble, France\\}

\begin{document}
\maketitle

\section{Introduction}

Generalised parton distributions (GPDs) have been extensively studied in various $ 2\to 2 $ scattering processes, such as deeply-virtual Compton scattering (DVCS) and deeply-virtual meson production (DVMP), where collinear factorisation has explicitly been shown to hold at all order in perturbation theory (in $  \alpha _{s} $) at the leading twist. It has been proposed that $ 2 \to 3 $ exclusive processes \cite{ElBeiyad:2010pji,Boussarie:2016qop,Pedrak:2017cpp,Duplancic:2018bum,Grocholski:2021man,Grocholski:2022rqj,Duplancic:2022ffo,Duplancic:2023kwe} would also admit a collinear factorisation, on the basis of the large invariant mass of the two particle produced in the reaction. In fact, the exclusive di-photon photoproduction process \cite{Grocholski:2021man,Grocholski:2022rqj} on a nucleon target is the only one that has been calculated at NLO in $  \alpha _{s} $. There, all IR singularities were shown to cancel. Quite recently, it was proved that collinear factorisation is valid for a family of $ 2 \to 3 $ exclusive processes \cite{Qiu:2022bpq,Qiu:2022pla}. It was pointed out that, rather than the large invariant mass of the produced pair of particles, it is the large relative transverse momentum between them that provided the hard scale to justify the collinear factorisation.

Nevertheless, when we computed the exclusive photoproduction of a $ \pi ^{0}\gamma   $ pair on a nucleon target, assuming collinear factorisation, we discovered that the amplitude for the gluon-induced channel (i.e.~with a gluon GPD) is \textit{divergent}, already at leading order and leading twist. The divergence occurred at the \textit{endpoint} of the outpoint pion distribution amplitude (DA), i.e.~when the momentum fraction $ z $ of one of the (anti)quark probed from the pion goes to zero, as well as at the \textit{breakpoint} of the GPD, i.e.~when the momentum fraction $ y $ of one of the gluons probed from the nucleon goes to zero. The term ``breakpoint'', rather than endpoint, emphasises the fact that the convolution of the hard partonic sub-process with the gluon GPD goes through $ y = 0 $, which gives rise to the so-called DGLAP and ERBL regions, determined by the sign of $ y $. While such endpoint/breakpoint regions have been shown not to cause any issues for the collinear factorisation of simpler processes like DVCS, this is clearly not the case for our process. The topic of these proceedings is precisely to investigate the origin of the divergence from this problematic region.

\section{Kinematics}

We denote the momenta of the particles in our process by
\begin{align}
\gamma (q) + N(p_N) \to \gamma' (q') + N'(p_{N'}) +  \pi ^{0}(p_{ \pi })\,
\end{align}
with $q^2 = q'^2 = 0$, $p_{N}^2 = p_{N'}^2= m_N^2$ and $ p_{ \pi }^{2} = m_{ \pi }^{2}$, 
where $ m_{N} $ and $ m_{ \pi } $ are the nucleon and pion mass respectively. We further define
\begin{align}
	P = \frac{p_{N}+p_{N'}}{2}\,,\qquad  \Delta  = p_{N'}-p_{N}\,,\qquad t =  \Delta ^{2}\,.
\end{align}
According to \cite{Qiu:2022pla}, collinear factorisation is expected to hold when $ q' $ and $ p_{ \pi } $ have large relative transverse momenta, $ |q'_{\perp}|,\, |p_{ \pi ,\,\perp}| $, in the centre-of-mass (CM) frame wrt to $  \Delta  $ and $ q $. This translates to the hard scale being  $ Q \sim |q'_{\perp}|,\,|p_{ \pi ,\,\perp}| $ with the requirement that $ Q $ is much larger than the soft scales in the process. Thus, defining the dimensionless parameter $  \lambda  $ given by
\begin{align}
	\label{eq:lambda-parameter}
	 \lambda  \sim \frac{\{\sqrt{|t|},\,m_{ \pi },\,m_{N},\, \Lambda _{ \mathrm{QCD} }\}}{Q}\,,
\end{align}
one expects a collinear factorisation of the process as $  \lambda  \to 0 $.

For our purposes, we find it more convenient to work in the CM frame wrt to $  \Delta  $ and $ p_{ \pi } $, especially for the power counting. This corresponds to having $  \Delta _{\perp} = p_{ \pi,\,\perp } = 0 $, with $ q $ and $ q' $ now having large transverse momentum. We thus work in a lightcone coordinate system where the momenta $ \Delta $ and  $ p_{ \pi } $ define the $ \bar{n} $ and $ n $ directions respectively. We represent a generic 4-vector $ V^{ \mu } $ by its lightcone components, $ V^{ \mu }  =  \left( V^{+},V^{-},V_{\perp} \right) $, with $ V^+ = n\cdot V $, $ V^{-} = \bar{n}  \cdot V $, and $ V_{\perp}^{\mu} = V^{ \mu } - V^{+} \bar{n}^{ \mu } - V^{-} n ^{ \mu } $, where $ \bar{n} $ and $ n $ are light-like vector that satisfy $ \bar{n}   \cdot n  = 1$. Thus, we obtain the following scalings for the components of the various momenta in the process,
\begin{align}
	p_{N},\,p_{N'},\, \Delta ,\,P \sim (1, \lambda ^2, \lambda )Q\,,\qquad
	p_{ \pi } \sim ( \lambda ^{2},1, \lambda )Q\,,\qquad
	q,\,q' \sim (1,1,1)Q\,,
\end{align}
with the condition that\footnote{The photons can also be \textit{quasi}-real, i.e.~$ q^2 \sim q'^2 \sim  \lambda ^{2} $. This has no effect on the arguments presented here.} $ q^{2} = q'^2 = 0 $. Since the photons have positive energy, this implies that $ q^{+},\,q^{-},\,q'^{+},\,q'^{-} > 0$. To simplify the notation, we fix $ Q =1 $ in what follows.

\section{Leading regions of partonic loop momenta}

It is convenient to study the amplitude $ {\cal A} $ for our process as a power series expansion in $  \lambda  $,
\begin{align}
	\label{eq:amplitude-expansion}
{\cal A} = \sum _{ \alpha } f_{ \alpha }  \lambda ^{ \alpha }\,.
\end{align}
Such an amplitude can be decomposed into various sub-graphs, such as soft $ S $, collinear $ C $ and hard $ H $, with internal partons of specific loop momenta connecting them together. As their names indicate, such sub-processes can only involve loop/external momenta that have specific scalings. For instance, a soft subgraph involves only soft momenta $ k_{s}  \sim  \left(  \lambda _{s}, \lambda _{s}, \lambda _{s} \right) $, where $  \lambda _{s}\ll 1 $. Note that $  \lambda _{s} $ is an arbitrary parameter which is independent of $  \lambda  $ fixed from the kinematics of our process, see \EQ\eqref{eq:lambda-parameter}. Of course, not any choice of $  \lambda _{s} $ are relevant in general. Typically, $  \lambda _{s} =  \lambda  $, which we refer to as the \textit{soft} scaling, and $  \lambda _{s} =  \lambda ^{2} $, which we call the \textit{ultrasoft} (or \textit{usoft}) scaling, are the only ones that matter.

The (u)soft scaling is to be distinguished from the \textit{Glauber scaling}, where
\begin{align}
	\label{eq:glauber-scaling}
k_{g}^{+}k^{-}_{g} \ll |k_{g,\,\perp}^{2}|\,,\quad \implies  k_{g} \sim ( \lambda ^2, \lambda ^2, \lambda ),\,( \lambda , \lambda ^{2}, \lambda ),\,...,
\end{align}
where there are different possibilities for how exactly the Glauber momentum $ k_{g} $ scales like. In particular, $ k_{g}\sim ( \lambda ^2, \lambda ^2, \lambda ) $ represents the more standard ``collinear-to-collinear'' Glauber scaling, which is well-known to be pinched in the classic Drell-Yan example \cite{Collins:2011zzd}. In our case, we find that it is the ``$ \bar{n} $-collinear-to-soft Glauber'' scaling, $ k_{g} \sim ( \lambda, \lambda ^2, \lambda ) $, that is relevant.

The aim then is to identify the leading (smallest $  \alpha  $) contribution in \EQ\eqref{eq:amplitude-expansion}, and to show that it can be decomposed in terms of simple \textit{universal} subgraphs, with only the hard subgraph $ H $ being process-dependent but calculable order-by-order in perturbative QCD. The statement that \textit{collinear factorisation} applies to a process at leading twist (or leading power) implies that there is a \textit{single} leading contribution to the amplitude, which involves only universal collinear subgraphs.\footnote{There can also be (u)soft subgraphs, but these typically give vacuum matrix elements of Wilson lines, and for many processes simply give unity.} The latter are the usual GPDs and DAs.
\begin{figure}
	\centering
	\hspace{-2.5cm}
	\begin{subfigure}{0.49\columnwidth}
		\includegraphics[width=4.5cm]{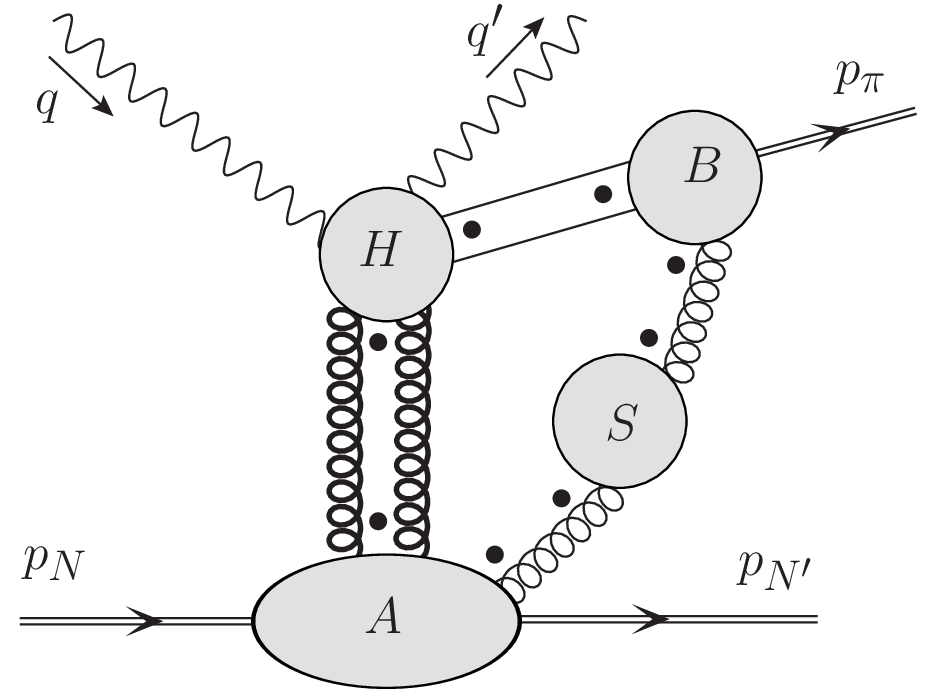}\\
		{\phantom{a}}\hspace{1.7cm}(a)\\[5pt]
		\includegraphics[width=4.5cm]{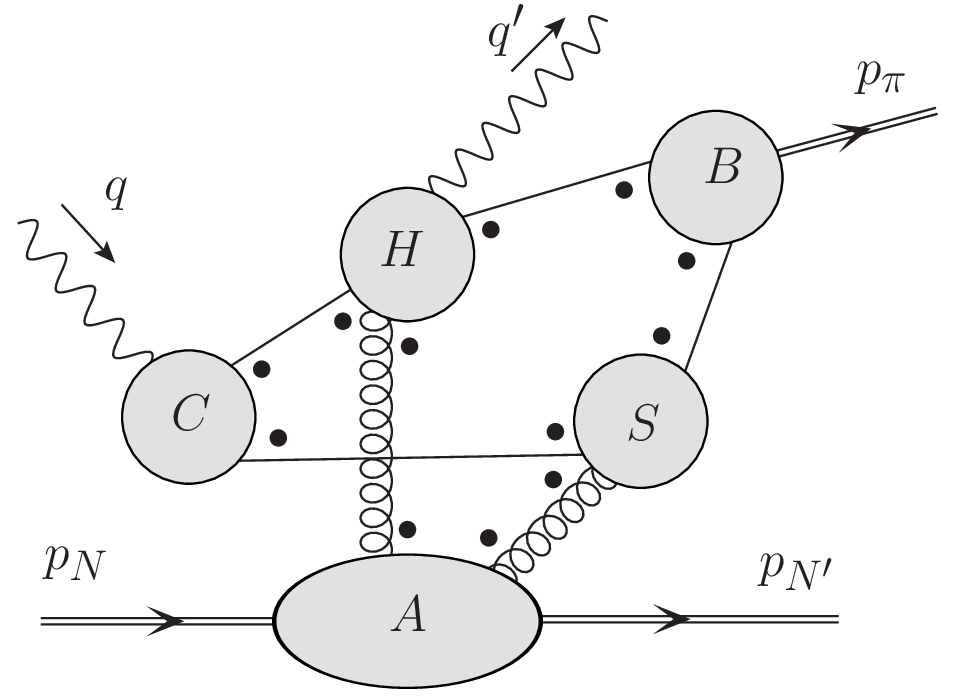}\\
		{\phantom{a}}\hspace{1.7cm}(b)
	\end{subfigure}
	\hspace{-2cm}
	\begin{subfigure}{0.49\columnwidth}
		\includegraphics[width=9cm]{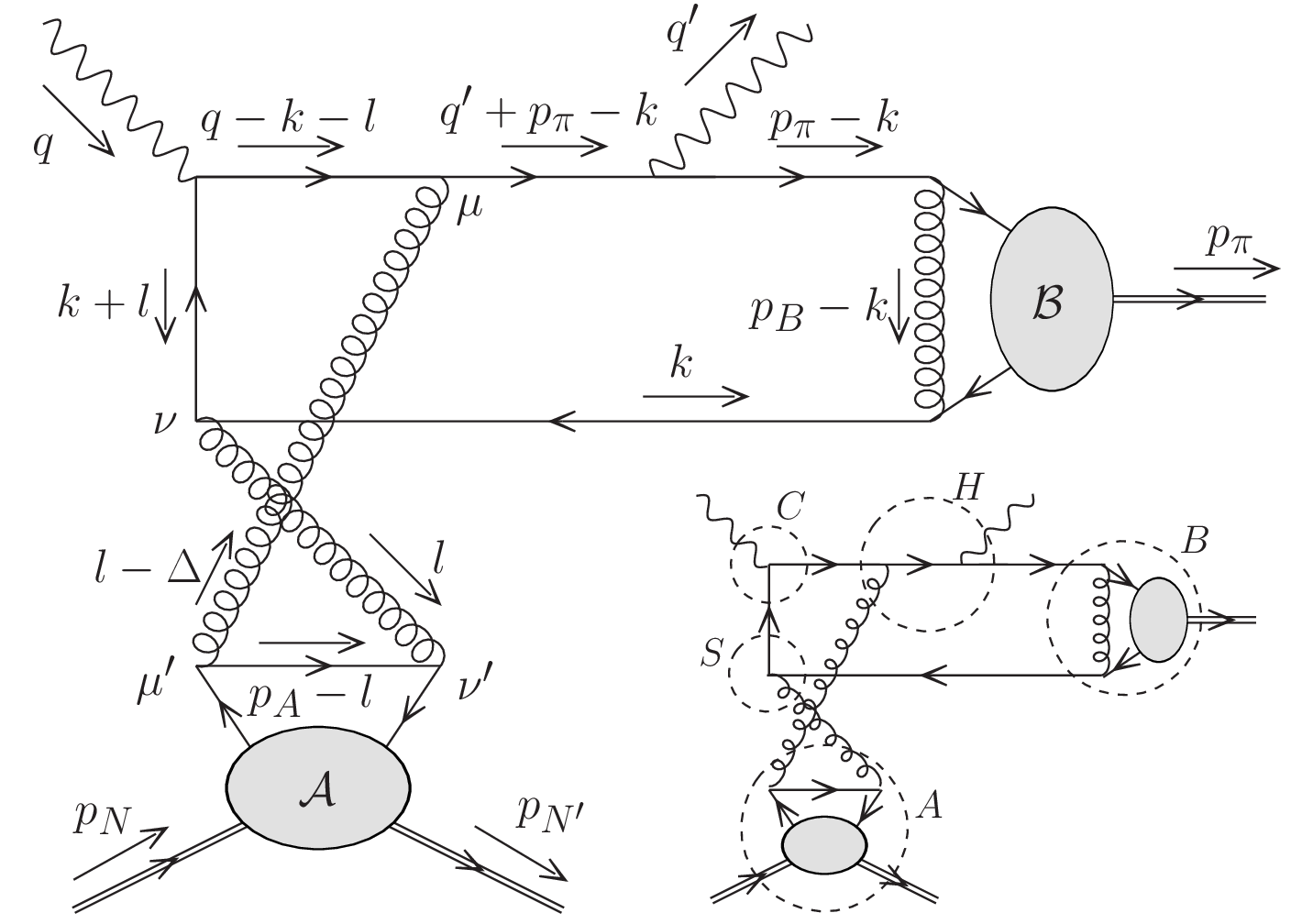}\\
			\end{subfigure}
	\caption{\textbf{Left}: Two relevant leading PSSs. $ A $ and $ B $ denote collinear regions for the incoming/outgoing nucleon and pion respectively. The dots next to the lines indicate an arbitrary number of longitudinally polarised gluons. (a): The collinear PSS. The thick gluon lines indicate that they are transversely polarised. (b): Another PSS involving the exchange of a soft gluon between the nucleon sector $ A $ and the soft quark line joining the pion $ B $ to the incoming photon $ C $, at $ S $.\\
		\textbf{Right}: An explicit 2-loop example to investigate the pinch structure and the power counting. For soft $ k $ and $ l $, it reduces to the PSS in (b), as shown on the bottom right.}
	\label{fig:reduced-diagrams-and-explicit-2-loop-example}
\end{figure}

The identification of the leading contribution and the associated subgraphs requires that the momenta of the internal partons connecting them to be \textit{pinched} to have the corresponding scaling. A loop momentum component is said to be pinched to be $ O( \lambda ) $, if there exist propagators with poles on opposite sides of the integration contour, with a separation of $ O( \lambda ) $. In the case of collinear pinches, this implies that the loop momentum connected to the collinear subgraph is forced to be, for instance, $ (1, \lambda ^{2}, \lambda ) $. The corresponding configuration of the loop momentum obtained by taking the limit $  \lambda \to 0 $ is called a \textit{pinch-singular surface} (PSS).

The Libby-Sterman analysis \cite{Libby:1978qf} allows the classification of pinched configurations and their associated power in $  \lambda  $.  Leaving the details to our main articles \cite{Nabeebaccus:2023rzr,Nabeebaccus:2024mia}, we simply point out here that the result of our analysis is that it suffices to focus on the two contributions in the left panel of \FIG\ref{fig:reduced-diagrams-and-explicit-2-loop-example}. On its own, \FIG\ref{fig:reduced-diagrams-and-explicit-2-loop-example}(a) would admit a collinear factorisation of our process. However, \FIG\ref{fig:reduced-diagrams-and-explicit-2-loop-example}(b) could ruin this picture, which involves the exchange of a soft gluon from the nucleon sector $ A $ to the soft quark line, passing through $ S $, which connects the incoming photon $ C $ to the outgoing pion $ B $. Strictly speaking, the Libby-Sterman analysis focuses only on cases with collinear, hard and usoft scalings. Therefore, in order to confirm that the diagram in \FIG\ref{fig:reduced-diagrams-and-explicit-2-loop-example}(b) is both pinched and of leading power when the soft gluon exchange has a collinear-to-soft Glauber scaling (see \EQ\eqref{eq:glauber-scaling}) and the quark through $ S $ has strictly soft scaling (i.e.~its momentum scales as $ ( \lambda , \lambda , \lambda ) $), one needs to demonstrate this explicitly. This is the subject of the next section.

\section{Explicit diagram with a leading power Glauber pinch}

Here, we focus on a particular fixed order  2-loop diagram (in $ l $ and $ k $) shown on the right panel of  \FIG\ref{fig:reduced-diagrams-and-explicit-2-loop-example}. In the Feynman gauge, the amplitude $ {\cal M} $ can be written as
\begin{align}
{\cal M} = \int F_{A}^{ \mu'  \nu '} g_{ \mu  \mu '} g_{ \nu  \nu '}  \mathrm{tr} \left[ F_{H}^{ \mu  \nu }F_{B} \right] \,,
\end{align}
where
\begin{align}
F_{A}^{ \mu ' \nu '} &= dl^{-}d^{2}l_{\perp}\frac{ \mathrm{tr} \left[ {\slashed{\cal A}} \gamma ^{ \nu'}  \left(  \slashed{p}_{A}-  \slashed{l}   \right)\gamma ^{ \mu '}   \right]  }{ \left[ l^2 +i \varepsilon \right]    \left[ \left( l- \Delta  \right)^2 +i \varepsilon \right]  \left[  \left( p_A - l \right)^2 + i \varepsilon   \right]  }\,,\\
F_{B}&= dk^{+}d^{2}k_{\perp}\frac{ \slashed{k} {\cal  \slashed{\cal B}  } \left(  \slashed{p}_{ \pi }- \slashed{k}   \right)   }{ \left[ k^2 + i \varepsilon  \right]   \left[ \left( p_{ \pi }-k \right)^2 + i \varepsilon  \right]   \left[  \left( p_{B}-k \right)^2 + i \varepsilon   \right]  }\,,\\
F_{H}^{ \mu  \nu } &= dk^{-}dl^{+}\frac{ \slashed{\epsilon }_{q'}^{*} \left(  \slashed{q}'+ \slashed{p}_{ \pi }- \slashed{k}    \right) \gamma ^{ \mu } \left(  \slashed{q}- \slashed{k}- \slashed{l}    \right) \slashed{\epsilon }_{q} \left(  \slashed{k}+ \slashed{l}   \right)  \gamma ^{ \nu }   }{  \left[ \left( q'+p_{ \pi }-k \right)^2+i \varepsilon  \right] \left[  \left( q-k-l \right)^2+i \varepsilon   \right]  \left[  \left( k+l \right)  ^2 + i \varepsilon \right]   }\,.
\end{align}
$ \epsilon _{q} $ and $ \epsilon _{q'}^{*} $ are the polarisation vector of the incoming and outgoing photons respectively. $ p_{A} $ and $ p_{B} $ are generic collinear loop momenta associated to their corresponding regions, such that $ p_{A} \sim  \left( 1, \lambda ^{2}, \lambda  \right)  $ and $ p_{B} \sim  \left(  \lambda ^{2},1, \lambda  \right)  $. $ {\cal A } $  and $  {\cal B }  $ are sub-amplitudes defined by their vector component in Dirac space,\footnote{For simplicity, we project quark anti-quark pair entering the pion onto the vector component in Dirac space. In reality, the projection is to be performed on the axial  vector component, but this has no consequence on our conclusions.} which include the external quark propagators as well as the measures $ d^{4}p_{A} $ and $ d^{4}p_{B} $ respectively. It can be argued that $ {\cal A } \sim  \left( 1, \lambda ^{2}, \lambda  \right) $ and $  {\cal B } \sim  \left(  \lambda ^{3},  \lambda  , \lambda ^{2} \right)   $ \cite{Nabeebaccus:2023rzr}.

\subsection{Glauber pinch}

We start from the result that the Landau equations predict that soft pinches are \textit{always} present. This is linked to the fact that a propagator with soft momentum, and its derivative, are both vanishing. Whether they are actually relevant and give a leading power contribution is a separate question. Thus, for the analysis of the Glauber pinch, we start from the soft scaling for the two loop momenta $ l $ and $ k $, i.e.~$ k,\,l \sim  \left(  \lambda , \lambda , \lambda  \right)  $, and determine whether the propagators force any of the lightcone components of  $ l $ to be much smaller.

A consideration of the propagators in $ F_{A}^{ \mu ' \nu '} $ leads to the pinching of $ l^{-} \sim  O( \lambda ^{2})  $,
\begin{align}
	 \left( l- \Delta  \right) ^{2} + i \varepsilon  =  -2 \Delta ^{+} \left(l^{-} +  O( \lambda ^{2}) + i \varepsilon  \right) \,,\quad
\left( p_{A}-l  \right) ^{2} + i \varepsilon  =  -2 p_{A} ^{+} \left(l^{-} +  O( \lambda ^{2}) - \sgn(p_{A}^{+})i \varepsilon  \right) \,,\nonumber
\end{align}
when $ p_{A}^{+}>0 $, noting that $  \Delta ^{+} < 0$ in our kinematics. The same mechanism leads to the pinching of the $ l^{-} $ component in the standard collinear pinch. At this point, we stress that the pinching of $ l^{-} $ alone does not imply that $ l $ is pinched in the Glauber region. This is because one may still be able to deform $ l^{+} $ to be $ O(1) $, such that $ l $ then becomes an $ \bar{n} $-collinear momentum. Such deformations from the Glauber scalings have been discussed in \cite{Collins:1997sr,Qiu:2022bpq}.

Looking at the poles in $ l^{+} $ from the propagators in $ F_{H} ^{ \mu  \nu }$, we find that
\begin{align}
 \left( k+l \right) ^2 + i \varepsilon  = 2k^{-}  \left( l^{+} + O( \lambda ) + \sgn(k^{-})i \varepsilon  \right) \,,\quad
 \left( q-k-l \right) ^2 + i \varepsilon  = -2q^{-}  \left( l^{+} + O( \lambda ) -i \varepsilon  \right) \,, \nonumber
\end{align}
so that $ l^{+} $ is pinched to be $ O( \lambda ) $ when $ k^{-}>0 $, noting that $ q^{-}>0 $. Therefore, we conclude that $ l $ is pinched to have $ \bar{n} $-collinear-to-soft Glauber scaling, $ l \sim  \left(  \lambda , \lambda ^{2}, \lambda  \right)  $.

\subsection{Power counting in the collinear region}

To determine the leading power from the collinear region, we perform the power counting on the amplitude $  {\cal M }  $, taking $ l \sim  \left( 1, \lambda ^{2}, \lambda  \right)  $ and $ k \sim  \left(  \lambda ^{2}, 1, \lambda  \right)  $. Furthermore, we take the all of the indices $  \mu,\, \nu ,\, \mu ',\, \nu ' $ to be \textit{transverse}, in order to obtain the ``correct'' leading power, and not a superleading power contribution which would eventually be suppressed through Ward identities (WIs). We refer the reader to \cite{Nabeebaccus:2023rzr, Nabeebaccus:2024mia} for a detailed discussion. Thus, we obtain
\begin{align}
F_{A}^{ \mu' _{\perp} \nu '_{\perp}} \sim  \lambda ^{4} \frac{ \lambda ^{2}}{ \lambda ^{6}} =  \lambda ^{0}\,,\qquad F_{B} \sim  \lambda ^{4}\frac{ \lambda ^{3}}{ \lambda ^{6}} =  \lambda \,,\qquad F_{H}^{ \mu _{\perp} \nu _{\perp}} \sim  \lambda ^{0}\frac{ \lambda ^{0}}{ \lambda ^{0}} =  \lambda ^{0}\,,
\end{align}
which fixes the leading power to be $  \lambda ^{1} $. We note that the same power is obtained through the Libby-Sterman power counting rules \cite{Libby:1978qf}.

\subsection{Power counting in the Glauber region}

Here, we take $ l \sim  \left(  \lambda , \lambda ^{2}, \lambda  \right)  $ and $ k \sim  \left(  \lambda , \lambda , \lambda  \right)  $. While the indices $  \mu , \mu ' $, which correspond to the $ \bar{n} $-collinear momentum $ l- \Delta  $ (see \FIG\ref{fig:reduced-diagrams-and-explicit-2-loop-example}), are taken to be transverse like in the above collinear power counting, it can be shown that choosing $  \nu  = \perp $ and $  \nu ' = + $ for the Glauber gluon gives the proper power counting for the diagram on the right panel of \FIG\ref{fig:reduced-diagrams-and-explicit-2-loop-example}, see  \cite{Nabeebaccus:2023rzr,Nabeebaccus:2024mia}. Thus, we get
\begin{align}
F_{A}^{ \mu' _{\perp} +} \sim  \lambda ^{4} \frac{ \lambda ^{1}}{ \lambda ^{6}} =  \lambda ^{-1}\,,\qquad F_{B} \sim  \lambda ^{3}\frac{ \lambda ^{3}}{ \lambda ^{4}} =  \lambda^{2} \,,\qquad F_{H}^{ \mu _{\perp} \nu _{\perp}} \sim  \lambda ^{2}\frac{ \lambda ^{1}}{ \lambda ^{3}} =  \lambda ^{0}\,,
\end{align}
such that the leading power is also $  \lambda ^{1} $, like the collinear region discussed above.

\section{Conclusion}

We have identified an $ \bar{n} $-collinear-to-soft Glauber pinch in the exclusive $  \pi ^{0}\gamma  $ pair photoproduction process, which we have also shown to contribute to the leading power like the standard collinear pinch. Furthermore, in \cite{Nabeebaccus:2023rzr, Nabeebaccus:2024mia}, we further explicitly show that the (u)soft pinch (which has the same topology as \FIG\ref{fig:reduced-diagrams-and-explicit-2-loop-example}(b)) is of subleading power. This means that the divergence in the amplitude, when collinear factorisation is assumed, is a direct consequence of the identified Glauber pinch in this paper, which breaks the collinear factorisation of the process. Moreover, such a Glauber pinch also occurs in similar $ 2 \to 3 $ exclusive processes that allow two-gluon exchanges in the $ t $-channel, such as the exclusive diphoton production from $  \pi ^{0}N $ scattering. Nevertheless, processes where only the quark exchange channel is present are safe from the factorisation breaking effects discussed here.

\small

\bibliographystyle{utphys}

\bibliography{masterrefs.bib}

\end{document}

%% file: useful-commands.tex
\def\slashchar#1{\setbox0=\hbox{$#1$}
	\dimen0=\wd0
	\setbox1=\hbox{/} \dimen1=\wd1
	\ifdim\dimen0>\dimen1
	\rlap{\hbox to \dimen0{\hfil/\hfil}}
	#1
	\else
	\rlap{\hbox to \dimen1{\hfil$#1$\hfil}}
	/
	\fi}

\newcommand{\FIG}{Fig.~}

\newcommand{\EQ}{Eq.~}

\def\bei{\begin{itemize}}
	\def\ei{\end{itemize}}

\def\beeq{\begin{eqnarray}} 
	\def\beqa{\begin{eqnarray}}
		\def\bea{\begin{eqnarray}}
			
			\def\eea{\end{eqnarray}}
		\def\eqa{\end{eqnarray}}
	\def\eeeq{\end{eqnarray}}

\def\eqar{\end{array}}
\def\beqar{\begin{array}}

\def\beas{\begin{eqnarray*}}
	\def\beqas{\begin{eqnarray*}}
		
		\def\eqas{\end{eqnarray*}}
	\def\eeas{\end{eqnarray*}}

\def\beq{\begin{equation}} 
	\def\be{\begin{equation}}
		
		\def\ee{\end{equation}}
	\def\eq{\end{equation}}
\def\eeq{\end{equation}}

\def\beqd{\begin{displaymath}}
\def\eeqd{\end{displaymath}}
\def\eqd{\end{displaymath}}

\def\beeq{\begin{eqnarray}} \def\eeeq{\end{eqnarray}}

\newcommand{\fin}{\end{document}}

\def\fin{\end{document}}

\DeclareMathOperator{\sgn}{sgn}